\documentclass{aa}
\usepackage{graphicx}
\input{psfig}
\begin{document}

\title{The $\zeta$\,Herculis binary system revisited}
\subtitle{Calibration and seismology}

\author{P. Morel, G. Berthomieu, J. Provost, F. Th\'evenin}

\institute{
D\'epartement Cassini, UMR CNRS 6529, Observatoire de la C\^ote 
d'Azur, BP 4229, 06304 Nice CEDEX 4, France.}

\offprints{P. Morel}
\mail{Pierre.Morel@obs-nice.fr}

\date{Received 9 July 2001 / Accepted 14 September 2001}

\abstract{ 
We have revisited the calibration of the 
visual binary system $\zeta$\,Herculis with the goal to
give the seismological properties of
the G0\,IV sub-giant $\zeta$\,Her\,A. The sum of masses and the mass fraction
are derived from the most recent astrometric data mostly based on the
{\sc hipparcos} ones. We have derived the effective temperatures,
the luminosities and the metallicities
from available spectroscopic data and {\sc tycho}
photometric data and calibrations. For the calculations of
evolutionary models we have used updated physics and the most recent physical
data. A $\chi^2$ minimization is performed to approach the most reliable
modeling parameters which reproduce the observations
within their error bars. For the age of the $\zeta$\,Her binary system we have
obtained $t_{\zeta\rm\,Her}=3\,387$\,Myr, for the masses
$m_{\zeta\rm\,Her\,A}=1.45\,M_\odot$ and $m_{\zeta\rm\,Her\,B}=0.98\,M_\odot$,
for the initial helium mass fraction $Y_{\rm i}=0.243$, for the initial mass
ratio of heavy elements to hydrogen $(\frac ZX)_{\rm i}=0.0269$ and for the
mixing-length parameters $\Lambda_{\zeta\rm\,Her\,A}=0.92$ and 
$\Lambda_{\zeta\rm\,Her\,B}=0.90$ using the Canuto \&
Mazitelli~(\cite{cm91}, \cite{cm92}) convection theory.
Our results do not exclude that $\zeta$\,Her\,A is itself
 a binary sub-system has been suspected many times in the past century;
the mass of the hypothetical unseen companion would be
$m_{\zeta\rm\,Her\,a}\loa 0.05\,M_\odot$, a value significantly smaller
than previous determinations.
A calibration made with an overshoot of the convective core of 
$\zeta$\,Her\,A leads to similar results but with a slight increase of $\approx
+250$\,Myr for the age.
The adiabatic oscillation spectrum of $\zeta$\,Her\,A is found to be a
complicated superposition of
acoustic and gravity modes. Some of these waves  have a dual character.
This greatly complicates the classification of the non-radial modes.
For $\ell=1$ the modes all have both 
energy in the core and in the envelope; they are mixed modes.
For $\ell=2,3$ 
there is a succession of modes with energy either in the core or in the 
envelope with a few mixed modes.
The echelle diagram used by the observers to extract the frequencies
will work for $\ell=0,\,2,\,3$. The large difference
is found to be of the order of $\overline{\Delta\nu_0}\approx 42\,\mu$Hz, in
agreement with the Marti\'c et al.~(\cite{mlsab01}) seismic observations.
}

\maketitle

\section{Introduction}\label{sec:int}
$\zeta$\,Herculis (40\,Her; BD\,+31\,2884; STF\,2084; ADS\,10157;
IDS\,16\,375\,+31\,47; WDS 16413+3136; HR\,6212; HD\,150680; HIP\,81693;
 $\alpha=16^{\rm h}\, 41^{\rm m}\,17^{\rm s}$,
$\delta=+31\degr36\arcmin10\arcsec\,(2000)$)
is a well known bright visual and single lined
spectroscopic binary system of naked-eye brightness.
According to the CDS Simbad data base
the system is composed of a 2.90 V\,magnitude G0IV sub-giant star and by a 5.53
V\,magnitude G7V dwarf star.
The binarity was discovered by
Herschell as early as 1782 (Aitken \cite{a32}) and the system has been carefully observed
 by visual binary observers for more than six revolutions
back to the first reliable measurements by
W. Struve in 1826. Several orbital solutions
have been published from the early nineties to the present day.
The latest are by Heintz~(\cite{h94}) and
S\"oderhjelm~(\cite{s00}). The orbital elements are well determined.
They have recently allowed improvements of
Sproul Observatory and {\sc hipparcos} trigonometrical parallaxes.
About a century ago, a duplicity of $\zeta$\,Her\,A was detected
from micrometer and meridian observations (Lewis~\cite{l06}).
A period $P_{\rm Aa}=12$\,yr and a semi-major axis of $a_{\rm Aa}=0\farcs25$
was obtained for the sub-binary system $\zeta$\,Her\,Aa. Ten years later,
Comstock~(\cite{c17}) noted that the small irregularities of the areal velocity
in the orbit have the effect of an invisible companion having a period
of $18$\,yr and an amplitude less than $0\farcs1$. Later,
a careful reanalysis of the available
astrometric and spectroscopic observational material led Berman~(\cite{b41}) to
conclude that{\it ``[..] the presence of a third body revolving about the brighter
component of $\zeta$\,Herculis is not definitely indicated''}. The comparison
between the observed position angles and distances and their values derived
from its astrometric orbit allowed Baize~(\cite{b76}) to claim that 
$\zeta$\,Her\,A is split in two stars of respectively $1.05\,M_\odot$ and
$0.19\,M_\odot$ with an orbital period $P_{\rm Aa}=10.5$\,yr and a
semi-major axis $a_{\rm Aa}=0\farcs06$.  Mc\,Carthy~(\cite{c83}) reported
 that $\zeta$\,Her contains at least one unseen companion easily detected at
 $2.2\,\mu$m by infrared speckle interferometry. 
A note in the Fifth Catalog of Orbits of Visual Binary Stars
(Hartkopf et al.~\cite{hmw00}) stipulates:
{\em ``No evidence in the speckle or Hipparcos data for the large-amplitude
third-body orbit given by Baize''}.

 Lebreton et al.~(\cite{lamb93}) tried to
determine both age and chemical composition of the system, modeling the
two components simultaneously, but they did not succeeded in modeling the secondary
consistently. Based on a precise spectroscopic analysis 
Chmielewski et al.~(\cite{cccls95}, hereafter C95)
succeeded in modeling both components. They
derived for the age $t_{\zeta\rm\,Her}=4.0\pm0.4$\,Gyr and for the masses of
components respectively $1.3\,M_\odot$ and $0.9\,M_\odot$.
Since 1995 the {\sc hipparcos}'s
parallax of $\zeta$\,Her has been available; it was recently improved by
S\"oderhjelm~(\cite{s00}). 
The {\sc tycho} \& {\sc hipparcos} magnitudes are also available
(Fabricius \& Makarov~\cite{fm00}); they
have been connected to the B, V and I magnitudes (Bessell~\cite{b00}). 
New improved theoretical data are also available, viz. opacities,
nuclear reaction rates and equation of state. The seismic observations of
$\zeta$\,Her\,A recently carried out by Marti\'c et al.~(\cite{mlsab01})
indicate the presence of solar like oscillations.

The aim of the present paper is firstly to
revisit the calibration of the $\zeta$\,Her binary system
using updated physical data and theories and improved astrometrical
and photometrical observational material, and secondly to give
seismological properties of $\zeta$\,Her\,A which will be useful
to exploit future asteroseismological observations.

Based on the reasonable hypothesis of a common origin for both components,
i.e. same initial chemical composition and age, the calibration of a binary
system consists of determining a consistent evolutionary history for the
double star, given (1) the positions of the two components in the {\sc hr} diagram,
(2) the stellar masses and, if possible, (3) the present-day surface chemical
compositions. The goal is to compute evolutionary models that reproduce the
observations.
The calibration yields estimates for the age, the
initial helium mass fraction and the initial metallicity
which are fundamental quantities for our
understanding of the galactic chemical evolution. Within the error bars
provided by astrometry, one also determines mass values 
consistent with the stellar structure modeling.
According to the convection theory applied, one derives
values for the ``mixing-length parameter'' or ``convection parameter''
$\Lambda$, ratio of the mixing-length to the pressure scale height. 
Once the physics is fixed, the modeling of the two components A \& B of a binary
system requires a set $\wp$ of seven so-called modeling parameters: 
\[\wp=\left\{t_\star ; m_{\rm A}, m_{\rm B}, Y_{\rm i},
\left(\frac ZX\right)_{\rm i}, \Lambda_{\rm A}, \Lambda_{\rm B}\right\},\]
where $t_\star$ is the age of the system, 
$m_{\rm A}$ and $m_{\rm B}$ are respectively the masses of components A \& B,
$Y_{\rm i}$ is the initial helium mass fraction, 
$\left(\frac ZX\right)_{\rm i}$ is the initial mass
ratio of heavy elements to hydrogen, $\Lambda_{\rm A}$ and
$\Lambda_{\rm B}$ are the mixing-length parameters.
There are six observables, namely, the effective
temperatures $T_{\rm eff\,A}$ \& $T_{\rm eff\,B}$, the
luminosities $L_{\rm A}$ \& $L_{\rm B}$,
the sum of masses $\cal S$ and the mass fraction $\cal B$ (cf. Eq.~\ref{eq:fm}).
Once detailed spectroscopic analyses have been performed on the
system, the present-day surface metallicities of stars
come as additional observational constraints
-- in the case of $\zeta$\,Her the metallicity is available
 only for the brightest component.
 
The paper is divided as follows. In Sect.~\ref{sec:astro}, \ref{sec:spectro}
and  \ref{sec:photo}, respectively, we collect and discuss the
astrometric, spectroscopic and photometric observations.
In Sect.~\ref{sec:comp}
we describe the computation of models and the search for the
modeling parameters.
In Sect.~\ref{sec:res}, we give the results with emphasis on the seismological
analysis of $\zeta$\,Her\,A. Section~\ref{sec:dis} is devoted to a discussion and
finally, we summarize our results and conclude in Sect.~\ref{sec:con}.

\section{Observations of the visual binary $\zeta$\,Her}
\begin{table}
\caption[]{ 
Relevant orbital elements of the $\zeta$\,Her binary system.
$P$ is the orbital period, $a$ the semi-major axis, $i$ the inclination and $e$
the eccentricity.
}\label{tab:dataa}
\begin{tabular}{llllllllllll} \hline \\
$P$     &$a$    &$i$   &$e$    &References\\
\\  \hline \\
$34.45$\,yr&$1\farcs365$&$131\fdg6$&$0.464$&Heintz~(\cite{h94})\\
$34.45$\,yr&$1\farcs33$ &$131\fdg$  &$0.46$ &S\"oderhjelm~(\cite{s00})\\
\\  \hline
\end{tabular}
\end{table}

\subsection{Astrometric data}\label{sec:astro}
The knowledge of individual masses of each companion is one
cornerstone of any calibration of a binary system. As illustrated by
calibrations of $\alpha$\,Cen, (e.g. Morel et al.~\cite{mpltb00}),
the resulting age is very sensitive to the values adopted
for the masses of the components. We do not reproduce here
the large bibliography on relevant astrometric data (e.g. C95); nevertheless
we emphasize the discussion concerning the estimate of masses.
For $\zeta$\,Her we are in the fortunate position of having two
precise and {\em independent} determinations of the
trigonometrical parallax and also improved orbital elements; from these
data, two estimates of the sum of masses can be derived independently.
A first determination is provided by a standard photographic
parallax (Heintz~\cite{h94}) using
an improved relative orbit and 152 mid-nights of several decades
 of Sproul Observatory long focus photographic observations. The
 second one is the recently improved adjustment of parallax and orbital
 elements by S\"oderhjelm~(\cite{s00})
based on {\sc hipparcos} data collected during 3.25\,yr, old ground based observations and recent
speckle-interferometry measurements.
Table~\ref{tab:dataa} lists the relevant orbital elements of these two
recent astrometric orbits; they are so close
that we can safely adopt the means listed in Table~\ref{tab:datad}.
The sum of masses is provided by Kepler's third law:
\begin{equation}\label{eq:sm}
{\cal S}\equiv m_{\rm A}+m_{\rm B}=\frac{a^3}{\varpi^3P^2},
\end{equation}
as usual $P$ (yr) is the period, $a$ (arc\,sec) is the
semi-major axis of the relative orbit of the
companion with respect to the primary,
$m_{\rm A}$ and $m_{\rm B}$ are the masses (solar unit)
of component A and B respectively and $\varpi$ (arc\,sec) is the parallax.

The outcome of standard long focus photographic parallax
measurements is the
so-called {\em relative} parallax $\varpi_{\rm rel}$. It must be reduced
to $\varpi_{\rm abs}$, the {\em absolute} parallax,
by adding a correction representing the dependence weighted parallax of
reference stars. The correction, of order 1 to 5 mas, is not accurately known
(van de\,Kamp~\cite{k67}).
For $\zeta$\,Her the correction is $0\farcs0043$ (Lippincott~\cite{l81}).
With $\varpi_{\rm rel}=0\farcs0974\pm0\farcs0039$ (Heintz~\cite{h94})
that leads to a photographic absolute parallax of
$\varpi_{\rm abs}=0\farcs102\pm0\farcs0039$\,(standard deviation). 
The improved absolute parallax based on {\sc hipparcos} data and orbital
ground-based measurements amounts to $\varpi_{\rm abs}=0\farcs0937\pm0\farcs0006$
(S\"oderhjelm~\cite{s00}).
We adopt respectively for the parallax and the distance modulus of $\zeta$\,Her:
\[\varpi_{\zeta\rm\,Her}=0\farcs094\pm0\farcs001,\ d_{\zeta\rm\,Her}=
-0.134\pm0.023.\]
The sum of masses calculated with Eq.~(\ref{eq:sm}) is:
\[{\cal S}_{\zeta\rm\,Her}=2.50\pm0.14\,M_\odot.\]
Note that the major source of error on
the sum of masses remains the uncertainty on $\varpi$
(Couteau~\cite{c78}, 40, VI) even with modern estimates of the parallax.
For the parallax, C95 have retained a value close to the former
{\sc hipparcos} result $\varpi_{\zeta\rm\,Her}=0\farcs0975$; with the same
orbital elements, this larger value for the parallax leads
to a smaller sum of masses ${\cal S}_{\zeta\rm\,Her}=2.4\,M_\odot.$

$\zeta$\,Her is one of the rare binary systems for which the mass fraction:
\begin{equation}\label{eq:fm}
{\cal B}=\frac{m_{\rm B}}{m_{\rm A}+m_{\rm B}},
\end{equation}
can be obtained both by
astrometric and spectroscopic observational data.
A first estimate of ${\cal B}$ is provided by photographic data.
As the orbits of the primary and secondary are similar with
respect to the center of masses, the semi-major axis $@$ of the orbit of the
primary is related to $a$ by $@={\cal B}a$.
For the so-called resolved binaries the two companions are distinguishable
by the detector and $@$ is directly derived from the observations.
For unresolved binaries, the distance between
the components is below the resolving power of the detector,
in which case no separation is possible, and a
composite image results. A close companion affects the center of light by
pulling it toward the barycenter. With the photographic technique
it is assumed that the measured position of
the blended image represents the weighted center of light-intensity, or
photocenter, of the components. In this case, the fractional distance $\beta$ of
the primary to the photocenter, in terms of the distance between the two
components, is given by (van\,de\,Kamp~\cite{k67}):
\begin{equation}\label{eq:phot}
\beta=\frac{L_{\rm B}}{L_{\rm A}+L_{\rm B}}=\frac1{1+10^{0.4\Delta m}}.
\end{equation}
Here $L_{\rm A}$ and $L_{\rm B}$ are the luminosities of the components and 
$\Delta m$ the difference in magnitude.
The fractional difference of photocenter
to barycenter is therefore ${\cal B}-\beta$ and the semi-major axis of the photocenter
orbit relative to the barycenter is $({\cal B}-\beta)a$. This $\beta$ correction also
explicitly affects the derivation of the parallax. It has
long been suspected
that for a magnitude difference $\Delta m \goa 1.5$ the
position of blended images generally cannot be
considered by the simple geometric Eq.~(\ref{eq:phot})
(Morel~\cite{m70}, Feierman~\cite{f71}). Likely, $\beta$ is also a function
of the separation on the photographic plate, among other parameters.
For the determination of the mass fraction of $\zeta$\,Her with $\Delta m=2.6$
(Hoffleit \& Jaschek~\cite{hj82})
Eq.~(\ref{eq:phot}) lead to $\beta=0.084$ while, from the work of
Feierman~(\cite{f71}), $\beta=0.025$ (Lippincott~\cite{l81}). These differences in
$\beta$ lead to the 
discrepancies between the mass fraction derived by Heintz,
${\cal B}=0.412\pm 0.020$, and Lippincott, ${\cal B}=0.352\pm 0.020$,
despite almost identical ${\cal B}-\beta$ values, namely 0.328 \& 0.32.
We shall refer to Heintz's (\cite{h94})
value of ${\cal B}_{\rm p}=0.41\pm0.02$ as the ``photographic'' mass ratio.

The second estimate of the mass fraction is provided by the spectroscopic orbit.
For a single lined spectroscopic binary, with known orbital elements and
parallax, the mass fraction is given by
(e.g. Heintz~\cite{h71}, Scarfe et al.~\cite{sfdb83}):
\begin{equation}\label{eq:b}
{\cal B}=\frac1{6.283}\frac{K_{\rm A}\varpi P\sqrt{1-e^2}}{a\sin i}.
\end{equation}
As usual $i$ is the inclination, $e$ is the
eccentricity and $K_{\rm A}$, in {\sc au}\,yr$^{-1}$, is the velocity semi-amplitude of the
primary star. Combined 
spectroscopic observations covering about four decades and astrometric
measurements give
$K_{\rm A}=4.01\pm0.04$\,km\,s$^{-1}$ (Scarfe et al.~\cite{sfdb83}).
From spectroscopy and orbital elements we obtain the ``spectroscopic''
mass fraction:
\[{\cal B}_{\rm s}=0.38\pm0.02.\]
We note the good agreement between spectroscopic and photographic
determinations.

\noindent
The mass fraction is not derived in the analysis of S\"oderhjelm~(\cite{s00})
despite an angular distance $\rho\goa1\farcs50$ larger than the grid step
($1\farcs208$) of {\sc hipparcos} (Martin et al.~\cite{mmf97}),
and changes of $\Delta\theta\sim10\degr$ and $\Delta\rho\sim0\farcs10$
 respectively
in position angle and separation along the flight of the satellite.

\noindent For further investigations we shall adopt as the mass fraction of 
$\zeta$\,Her the weighted mean:
\[{\cal B}_{\zeta\rm\,Her}=0.40\pm0.02,\]
between astrometric and spectroscopic values.
C95 have adopted a slightly larger value ${\cal B}_{\zeta\rm\,Her}=0.42$.

\noindent With the values retained for 
${\cal S}_{\zeta\rm\,Her}$ and ${\cal B}_{\zeta\rm\,Her}$ 
the individual masses are:
\[m_{\zeta\rm\,Her\,A}=1.50\pm0.16, \quad m_{\zeta\rm\,Her\,B}=1.00\pm0.08.\]
Table~\ref{tab:datad} lists the astrometric data and the astrometric
constraints we use for the calibration of the $\zeta$\,Her binary system.

\subsection{Spectroscopic data}\label{sec:spectro}
Effective temperatures and luminosities can be determined from both photometric
and spectroscopic analyses.
As a general rule, when available, it is the detailed spectroscopic analysis,
using hydrogen line profiles, which provides the best estimates for effective
temperatures, while bolometric magnitudes and luminosities
are more accurately determined using photometric data and calibrations.

With classical 1.5\,m to 2.0\,m telescopes $\zeta$\,Her appears as a single star
under average seeing conditions because of the large magnitude
difference and of the small angular distance. Therefore
isolated spectra of each component cannot be obtained.
The available spectroscopic data in C95
 allows the derivation of effective temperature only for
$\zeta$\,Her\,A. We start with an estimate of $T_{\rm eff\,A}$ from
 Magain's~(\cite{m87}) calibration, assuming a metallicity of [Fe/H]=0.0
and color index (B-V)=0.65 (see Table~\ref{tab:datap}).
Then we perform a standard {\sc lte} detailed analysis using the
curve of growth technique with models atmosphere from Gustafsson
et al.~(\cite{gben75}). Equivalent widths used are from C95 and
the oscillator strengths are from Th\'evenin~(\cite{t90}).
The microturbulence is fixed to $1.5$\,km\,s$^{-1}$.
With the updated solar iron abundance $\log\epsilon_{\rm Fe}=7.46$
(Holweger~\cite{h79}, Asplund~\cite{a00}) we
derived the same metallicity [Fe/H]=0.04 dex as C95. An ionization 
equilibrium is obtained from curves of growth of FeI and FeII which permits us to
derive the surface gravity $\log g=3.85$.
We correct the [Fe/H] value in
Magain's~(\cite{m87}) formula and derived an improved temperature
value of $T_{\rm eff\,A}=5\,820\pm50$\,K and re-iterate the curve of growth
analysis.
Finally we deduce $\log g_{\rm A}=3.75\pm0.15$ and $\rm [Fe/H]_A=+0.04\pm0.03$.
With these improved input parameters
the scattering of the curve of growth decreases and is satisfactory.
 C95 have derived a slightly smaller value for
 the gravity $\log g_{\rm A}=3.65\pm0.20$ invoking non-{\sc lte} effects.
Such stars with solar abundances are not suspected to suffer from {\sc nlte}
overionization (Th\'evenin \& Idiart~\cite{ti99}), therefore the surface
gravity of $\zeta$\,Her\,A can be considered as being well determined.
It is remarkable that this new effective temperature value we derived,
using both {\sc tycho}'s photometry (Fabricius \& Makarov~\cite{fm00})
and spectroscopic analysis, is very close to $5\,825\pm40$\,K, the value
obtained by C95 using both the
photometry available in 1995 and the profile of
H$\alpha$ corrected by the presence of the light of the secondary.
The uncertainty on $\rm [Fe/H]_A$ has been estimated by varying 
$T_{\rm eff}$ and $\log g$
in their extremes and with the accuracy of the fit resulting from
the quality of equivalent widths.
The precision obtained for the metallicity is compatible with a value of
$0.05$\,dex, which is the more pessimistic estimate in C95. 
The precision obtained for the effective temperature results from
the partial derivative of Magain's~(\cite{m87}) temperature
formula combined with the abundance uncertainty and errors of the 
{\sc tycho} magnitudes listed Table~{\ref{tab:datap}. For the effective
temperature of $\zeta$\,Her\,A, C95 have obtained $\pm 40$\,K,
a similar estimate.

The projected rotational velocity $v\sin i$ of $\zeta$\,Her\,A amounts
to $3.9$\,km\,s$^{-1}$ (Fekel~\cite{f97}). Assuming a parallel axis for
 rotation and orbital motion, $\sin i\approx0.655$,
$\zeta$\,Her\,A is therefore a slow rotator. We can safely infer that
 it is the same for the less massive B component. So we
can neglect the small rotational velocity in modeling the internal
structure of both components.

\subsection{Photometric data}\label{sec:photo}
We base the photometric analysis on magnitudes and color indexes
taken from the {\sc tycho} catalogue (Fabricius \& Makarov~\cite{fm00})
as it provides consistent data for both components. For each component
{\sc tycho}'s
photometric system gives magnitudes $\rm V_T$ and $\rm B_T$ with an accuracy of
the order of $0.01$\,mag. {\sc hipparcos}'s photometric system gives 
$\rm H_P$ magnitudes with an accuracy of
the order of a few milli-magnitude. We use Bessell's~(\cite{b00})
relationships between color index $\rm (B_T-V_T)$ and $\rm V-H_P$, to derive the  
magnitudes in standard Johnson visual $\rm V$ and blue $\rm B$ filters.
In view of figures 3 and 4 of Bessell's~(\cite{b00}) paper,
the accuracy of calibrations are estimated to $\pm0.03$\,mag.
The bolometric corrections, derived from Bessell et
al.~(\cite{bcp98}) for each star are respectively:
\[{\rm BC_V(A)}=-0.0566\pm0.008, \quad {\rm BC_V(B)}=-0.168\pm0.038.\]
To be consistent with the detailed
spectroscopic analysis of $\zeta$\,Her\,A, 
the effective temperature of $\zeta$\,Her\,B is derived from
Magain's~(\cite{m87}) formula using the color index $\rm (B-V)$
listed Table~\ref{tab:datap}, the metallicity derived for
$\zeta$\,Her\,A and the standard gravity of dwarf
stars $\log g\sim 4.44$.
The accuracy on $T_{\rm eff\,B}$ is derived as for
$T_{\rm eff\,A}$. Resulting from the high accuracy of {\sc tycho} magnitudes, we
derive a better precision than C95. We obtain
 $T_{\rm eff\,B}=5\,300\pm150$\,K which is close to $5\,290\pm300$\,K,
 the value obtained by C95.

Table~\ref{tab:datap} lists the bolometric corrections, the
bolometric magnitudes and luminosities computed
with $M_{\rm bol\,\odot}=4.74$ as the solar bolometric magnitude to be used with
the Bessell et al.~(\cite{bcp98}) calibration.
 The Bessell~(\cite{b00}) photometric calibration allows us to derive
the standard Johnson B and V magnitudes
either from $(\rm B_T \& V_T)$ {\sc tycho} magnitudes, or from
$(\rm B_T \& V_T)$ and $\rm H_P$ {\sc hipparcos} magnitudes.
The values obtained from each differ from each other by more than is
expected from the accuracy of measurements. Moreover, fixing the color index
(B-V), the effective temperatures calculated with different
photometric calibrations (Magain~\cite{m87},
Alonso et al.~\cite{aam96}, Flower~\cite{f96}) differ by more than 150\,K.
This indicates that the errors on data derived with the whole procedure
may not have a standard Gaussian distribution.
Therefore the uncertainties of quantities derived from photometric data
are added in absolute value (${\cal L}^1$ norm) instead of standard quadrature 
(${\cal L}^2$ norm).
Table~\ref{tab:datad} allows comparisons between
the observational constraints
 retained in C95 and in this paper. The main
 differences are for the luminosity values owing to the smaller
 bolometric corrections and parallax we have used.

\begin{table}
\caption[]{
Photometric data from {\sc tycho} and {\sc hipparcos} and
derived bolometric corrections, magnitudes and luminosities.
}\label{tab:datap}
\begin{tabular}{llllllllllll} \hline \\
&$\zeta$\,Her\,A                   &$\zeta$\,Her\,B \\
\\ \hline \\
$\rm B_T$ &$3.70\pm0.01$       &$6.39\pm0.01$\\
$\rm V_T$ &$2.98\pm0.01$       &$5.47\pm0.01$\\
$\rm H_p$ &$3.022\pm0.002$     &$5.705\pm0.027$\\
\\
$\rm B$   &$3.539\pm0.055$     &$6.369\pm0.081$\\
$\rm V$   &$2.890\pm0.033$     &$5.556\pm0.059$\\
$\rm (B-V)$ &$0.649\pm0.088$    &$0.81\pm0.140$\\
\\
$\rm BC_V$  &$-0.0566\pm0.009$   &$-0.168\pm0.026$\\
$M_{\rm bol}$  &$2.699\pm0.065$   &$5.254\pm0.108$\\
$L/L_\odot$ &$6.55\pm0.39$        &$0.62\pm0.06$ \\
\\  \hline
\end{tabular}
\end{table}

\begin{table}
\caption[]{ 
Adopted observational data and derived constraints for the calibration of
$\zeta$\,Her binary system.
}\label{tab:datad}
\begin{tabular}{llllllllllll} \hline \\
& this paper & Chmielewski et al.~(\cite{cccls95})\\
\\ \hline \\
$P$                        &$34.45\pm0.005$\,yr         & \\        
$a$                        &$1\farcs35\pm0\farcs02$     & \\ 
$i$                        &$131\fdg3\pm0\fdg3$         & \\
$e$                        &$0.462\pm0.002$             & \\
$K_{\rm A}$                &$4.010\pm0.04$\,km\,s$^{-1} $& \\
$\varpi$                   &$0\farcs094\pm0\farcs001$   &$0\farcs0975$\\
\\
${\cal S}_{\zeta\rm\,Her}$ &$2.50\pm0.14\,M_\odot$      &$2.40\,M_\odot$ \\
${\cal B}_{\zeta\rm\,Her}$ &$0.40\pm 0.02$              &$0.42$ \\
$T_{\rm eff\,A}$           &$5820\pm50$\,K              &$5825\pm40$\,K \\
$L_{\rm A}/L_\odot$        &$6.55\pm0.39$               &$6.194\pm0.285$ \\
$\rm [\frac{Fe}H]_{\rm A}$ &$0.04\pm0.03$               &$0.05\pm0.05$ \\
$T_{\rm eff\,B}$           &$5300\pm150$\,K             &$5290\pm300$\,K \\
$L_{\rm B}/L_\odot$        &$0.62\pm0.06$               &$0.575\pm0.069$ \\
\\  \hline
\end{tabular}
\end{table}

\section{Evolutionary models}\label{sec:comp}
The mean angular separation between $\zeta$\,Her\,A \& B
being of the order of
$\bar\rho=1\farcs5$, the distance between the two companions
amounts to $\simeq150$\,{\sc au}. The tidal interactions between
the two components are then negligible. We can safely assume that stars
 have evolved as single ones, ignoring each other.

Models have been computed using the {\sc cesam} code (Morel~\cite{m97}).
About 600 mass shells describe each model; this number
 increases up to 2100 for the model used in seismological analysis.
Around 400 and 30 models are needed to describe the evolutions of
$\zeta$\,Her A and B respectively.

Basically the physics employed is the same as in
Morel et al.~(\cite{mpltb00}). The ordinary assumptions of stellar modeling are
made, i.e. spherical symmetry, no rotation, no magnetic field and no mass loss.
The evolutions are initialized with homogeneous zero-age main-sequence models
({\sc zams}).
In the absence of satisfactory treatment of microscopic diffusion for
stars with mass larger than $\approx 1.3\,M_\odot$, we do not
take into account the diffusion of chemical species. This important
assumption is discussed in Sect.~\ref{sec:dis}.

\subsection{Physical inputs}
The relevant nuclear reaction rates
are taken from the {\sc nacre} compilation (Angulo et al.~\cite{aar99}).
We use the approximation:
\begin{eqnarray}\label{eq:fesh}
\rm{[\frac{Fe}H]} &\equiv&\log(\frac{Z_{\rm Fe}}Z)+\log\left(\frac ZX\right)-
\log(\frac{Z_{\rm Fe}}X)_\odot\simeq \nonumber \\
&\simeq&\log\left(\frac ZX\right) - \log\left(\frac ZX\right)_\odot \nonumber
\end{eqnarray}
where $(\frac{Z_{\rm Fe}}Z)$ is the iron mass fraction within $Z$. We 
use the solar mixture of Grevesse \& Noels~(\cite{gn93}) i.e.
$\left(\frac ZX\right)_\odot=0.0245$.
We employed the {\sc opal} equation of state (Rogers et al. \cite{rsi96})
 and the opacities of Iglesias \& Rogers~(\cite{irw92})
complemented at low temperatures by
Alexander \& Ferguson~(\cite{af94}) opacities.
In the convection zones the temperature gradient is
computed according to the Canuto \& Mazitelli~(\cite{cm91}, \cite{cm92})
convection theory.
The mixing-length is defined as $l\equiv \Lambda H_{\rm p}$,
where $H_{\rm p}$ is the pressure scale height and $\Lambda$ is
the mixing-length parameter of order unity. $\zeta$\,Her\,A,
more massive than $1.25\,M_\odot$, presents a
convective core along the main-sequence. Following
the prescriptions of Schaller et al.~(\cite{ssmm92}) we have also calibrated the
$\zeta$\,Her binary system with an overshooting of the 
convective core of $\zeta$\,Her\,A over the distance
$O_{\rm v}=0.2\min(H_{\rm p}, R_{\rm co})$,
where  $R_{\rm co}$ is the core radius.
The atmosphere is restored using
Hopf's law (Mihalas~\cite{m78}).
We use the numerical trick of Henyey et al.~(\cite{hvb65}) to
connect consistently the radiative and the convective parts of the atmosphere.
Hence a smooth connection of the
gradients is insured between the uppermost layers of the envelope
and the optically thick convective bottom of the atmosphere.
It is an important requirement for the
calculation of eigenmode frequencies.
The radius $R_\star$ of any model
is taken at the optical depth $\tau_\star$ where
$T(\tau_\star)=T_{\rm eff}$, $\tau_\star=0.645$ with Hopf's law.
The mass $M_\star$ of the star is defined as the mass enclosed in
the sphere of radius $R_\star$.
The external boundary is located at the optical
depth $\tau_{\rm ext}=10^{-4}$, where the density is fixed at a standard
 value in Kurucz's~(\cite{k91}) atmosphere models:
 $\rho(\tau_{\rm ext})=3.55\,10^{-9}$\,g\,cm$^{-3}$.

\subsection{The search of calibration parameters}\label{sec:chi2}
The calibration of a binary system is based on the adjustment
of stellar modeling parameters
to observational data at the age of the system.  Fixing the physics,
the effective temperature $T_{\rm eff}$, the luminosity $L$ and the surface
metallicity $\rm\left[\frac{Fe}H\right]_s$ of
a stellar model have the formal dependences with respect to modeling parameters:
\begin{eqnarray}
&&\log T_{\rm eff}(\star)_{\rm mod}=\log T_{\rm eff}
\left(t_\star;m_\star,Y_{\rm i\star},{\rm[\frac{Fe}H]_i\star},\Lambda_\star\right),\nonumber\\
&&\log \left(\frac{L(\star)}{L_\odot}\right)_{\rm mod}=\log\left(\frac L{L_\odot}\right)
\left(t_\star;m_\star,Y_{\rm i\star},
{\rm[\frac{Fe}H]_i\star},\Lambda_\star\right),\nonumber \\
&&{\rm[\frac{Fe}H]_{\rm s}}(\star)_{\rm mod}=
{\rm[\frac{Fe}H]_{\rm s}}\left(t_\star;m_\star,Y_{\rm i\star},
{\rm[\frac{Fe}H]_i\star},\Lambda_\star\right), \label{eq:fe}
\end{eqnarray}
where the subscript ``$\rm _{mod}$'' refers to model values.
Without microscopic diffusion almost no change in chemical composition due to
nuclear reactions occurs in the envelope of a star as massive as $\zeta$\,Her\,A
so $\rm[\frac{Fe}H]_{\rm s}\approx[\frac{Fe}H]_{\rm i}$ and
Eq.~\ref{eq:fe} becomes trivial.
The basic idea of the $\chi^2$ fitting has been developed by
Lastennet et al.~(\cite{lvlo99}). To find a set of modeling parameters: 
\[\wp_{\zeta\rm\,Her} \equiv\left\{t,m_{\rm A},m_{\rm B},
Y_{\rm i},\left(\frac ZX\right)_{\rm i},\Lambda_{\rm A},\Lambda_{\rm B}\right\}_{\zeta\rm\,Her},\]
leading to observables as close as possible to the observational constraints
$\log T_{\rm eff}(\zeta{\rm\,Her\,A})$,
$\log \left(\frac{L(\zeta{\rm\,Her\,A})}{L_\odot}\right)$,
$\log T_{\rm eff}(\zeta{\rm\,Her\,B})$,
$\log \left(\frac{L(\zeta{\rm\,Her\,B})}{L_\odot}\right)$ and
$\rm[\frac{Fe}H]_{\rm s}(\zeta{\rm\,Her\,A})$
we minimize a
$\chi^2(t_\star,m_{\rm A},m_{\rm B},Y_{\rm i},{\rm[\frac{Fe}H]_i},
\Lambda_{\rm A},\Lambda_{\rm B})$ functional defined as:
\begin{eqnarray}
\chi^2&=&
\left(\frac{{\rm [\frac{Fe}H]_s}(A)_{\rm mod}-{\rm [\frac{Fe}H]_{\rm s}}(A)}
{\sigma\left({\rm [\frac{Fe}H]_{\rm s}}(A)\right)}\right)^2 +\nonumber \\
&+&\sum_{\star=A,B}\left[\left(\frac{\log T_{\rm eff}(\star)_{\rm mod}-
\log T_{\rm eff}(\star)}
{\sigma\left(\log T_{\rm eff}(\star)\right)}\right)^2 + \right. \nonumber\\
&+&\left.\left(\frac{\log\left(\frac{L(\star)}{L_\odot}\right)_{\rm mod}
-\log\left(\frac{L(\star)}{L_\odot}\right)}
{\sigma\left(\log\left(\frac{L(\star)}{L_\odot}\right)\right)}\right)^2\right]
 \label{eq:chi2}
\end{eqnarray}
where the $\sigma$'s are the uncertainties associated with the constraints.
For sets of modeling parameters within the ranges detailed
in Sect.~\ref{sec:cal}, we have computed the evolution of models
from homogeneous {\sc zams} ($t=0$) to $t\loa4$\,Gyr.
Then the $\chi^2$ was computed using Eq.~(\ref{eq:chi2}) in a refined grid 
obtained by interpolations. We kept for the solution
the ``best'' $\wp=\wp_{\zeta\rm\,Her}$ which corresponds
to the $\chi^2_{\rm min}$. After some hand adjustments
these best modeling parameters served to compute the
 models of $\zeta$\,Her\,A \& B.
Table~\ref{tab:glob} lists the confidence limits of the modeling parameters
of models computed in this paper.
The confidence limits of each modeling parameter, the other being fixed, 
correspond to the maximum/minimum values it can reach, in order that the
generated models fit the observable targets within their error bars. 
Owing to the uncertainties of the observation material, mainly on the effective
temperature of $\zeta$\,Her\,B, we do not attempt
to improve the solution, or to compute more precisely the
error bars affecting the modeling parameters using the singular value
decomposition of the {\em design matrix} as developed by Brown et
al.~(\cite{bcwg94}).

\section{Results}\label{sec:res}
\subsection{Calibration of $\zeta$\,Her}\label{sec:cal}
We have computed the evolution of models with
initial helium mass fraction, initial ratio of heavy elements to hydrogen,
masses and mixing length parameters respectively in the ranges
$0.24\leq Y_{\rm i}\leq 0.27$,
$0.026\leq [\frac ZX]_{\rm i}\leq 0.030$,
$1.35\,M_\odot \leq m_{\zeta\rm\,Her\,A}\leq 1.65\,M_\odot$, 
$0.85\,M_\odot \leq m_{\zeta\rm\,Her\,B}\leq 1.15\,M_\odot$,
$0.9 \leq \Lambda_{\rm\zeta\,Her\,A}\leq 1.1$ and
$0.9 \leq \Lambda_{\rm\zeta\,Her\,B}\leq 1.1$.
For the models of $\zeta$\,Her\,A, as soon as
$\log T_{\rm eff\,A}\leq 3.765$ the evolution is halted.
Table~{\ref{tab:glob} lists the set of ``best'' modeling parameters
we adopt for the $\zeta$\,Her A \& B models
with and without overshooting of the convective
core of the primary. Figure~\ref{fig:AB} shows the corresponding
evolutionary tracks in the {\sc hr} diagram. For the
calibration with core overshooting of $\zeta$\,Her\,A,
we have not recomputed specific sets of models
to undertake another $\chi^2$ minimization. We fit the
observable constraints within error boxes by adjustments of
$t_\star$ and $(\frac ZX)_{\rm i}$,
the other modeling parameters having the values obtained without
overshooting. The larger value obtained for the age ($\sim+250$\,Myr)
results from the larger
amount of nuclear fuel available in the overshoot convection core.

\begin{table}
\caption[]{
Calibration parameters of $\zeta$\,Her binary system models
lying within the uncertainty
boxes. The observational constraints listed in Table~\ref{tab:datad}
are recalled in parenthesis.
The subscripts ``$_{\rm i}$'' and ``$_{\rm s}$'' respectively refer
to initial and surface
quantities.
}\label{tab:glob}
\begin{tabular}{cccc} \\  \hline \\
          &$\zeta\,\rm Her\,A$  &  $\zeta\,\rm Her\,B$ \\  \\ \hline \\
$m/M_\odot$         &$1.45\pm 0.01$ &$0.98\pm 0.02$\\
${\cal S}\,(M_\odot)$&\multicolumn{2}{c}{$2.43\,(2.50\pm0.14)$} \\
${\cal B}$&\multicolumn{2}{c}{$0.403\,(0.40\pm0.02)$} \\
$Y_{\rm i}$ &\multicolumn{2}{c}{$0.243\pm 0.002$}  \\ 
$\rm [\frac{Fe}H]_{\rm s}$ &$0.042 \,(0.04\pm0.003)$  \\
$\Lambda$ &$0.92\pm0.05 $&$0.90\pm0.10 $ \\
 \\ \hline \\ 
 &\multicolumn{2}{c}{Calibration without overshoot for $\zeta$\,Her\,A} \\ \\
age (Myr) &\multicolumn{2}{c}{$3\,387\pm15$} \\
$T_{\rm eff}$\,(K)   &$5\,818\,(5\,820\,\pm50)$ & $5\,413\,(5\,300\,\pm150)$ \\
$L/L_\odot$      &$6.72\,(6.55\pm0.39)$ & $0.65\,(0.62\pm0.06)$ \\      
$\left(\frac ZX\right)_{\rm i}$ &\multicolumn{2}{c}{$0.0269\pm 0.0005$} \\       
$X_{\rm i}$ &\multicolumn{2}{c}{$0.737 $}  \\
$Z_{\rm i}$ &\multicolumn{2}{c}{$0.0198$}  \\                  
$ R/R_\odot$         & $2.56$ &$ 0.915$ \\
$\overline{\Delta\nu_0}\,(\mu$Hz) & $42.2$ & $157.$\\
 \\ \hline \\
 &\multicolumn{2}{c}{Calibration with overshoot for $\zeta$\,Her\,A} \\ \\
age (Myr) &\multicolumn{2}{c}{$3\,625\pm22$}\\
$T_{\rm eff}$\,(K)   &$5\,798\,(5\,820\,\pm50)$ & $5\,418\,(5\,300\,\pm150)$ \\
$L/L_\odot$      &$6.82\,(6.55\pm0.39)$ & $0.65\,(0.62\pm0.06)$\\
$\left(\frac ZX\right)_{\rm i}$ &\multicolumn{2}{c}{$0.0272\pm 0.005$} \\ 
$X_{\rm i}$ &\multicolumn{2}{c}{$0.737$} \\
$Z_{\rm i}$ &\multicolumn{2}{c}{$0.0200$} \\                  
$ R/R_\odot$         & $2.61$ &0$.920$ \\
$\overline{\Delta\nu_0}\,(\mu$Hz) & $40.9$ & $156.$\\
 \\ \hline
\end{tabular}
\end{table}

\begin{table}
\caption[]{
Mass ($M_\odot$), initial mass fraction of helium, initial ratio
of heavy elements to hydrogen, age (Myr), radius ($R_\odot$),
effective temperature (K) and large difference ($\mu$\,Hz)
of $\zeta\,\rm Her\,A$ models -- without overshooting --
 computed with modeling parameters
within the error bar of Table~\ref{tab:glob}.
}\label{tab:acc}
\begin{tabular}{llllllll} \\  \hline \\
 & $M$ & $Y_{\rm i}$ & $(\frac ZX)_{\rm i}$ &  age & $R$ & $T_{\rm eff}$ & $\overline{\Delta\nu_0}$ \\ 
\\ \hline \\
Ag$+$ & $1.45$ & $0.243$ & $0.0269$ & $3400$ & $2.58$ & $5763$ & $41.6$\\
Ag$-$ & $1.45$ & $0.243$ & $0.0269$ & $3372$ & $2.53$ & $5868$ & $42.8$\\
Am$+$ & $1.46$ & $0.243$ & $0.0269$ & $3315$ & $2.59$ & $5811$ & $41.6$\\
Am$-$ & $1.44$ & $0.243$ & $0.0269$ & $3415$ & $2.53$ & $5808$ & $42.7$\\
Ay$+$ & $1.45$ & $0.245$ & $0.0269$ & $3298$ & $2.57$ & $5808$ & $41.8$\\
Ay$-$ & $1.45$ & $0.242$ & $0.0269$ & $3332$ & $2.55$ & $5811$ & $42.3$\\
Az$+$ & $1.45$ & $0.243$ & $0.0274$ & $3376$ & $2.55$ & $5810$ & $42.4$\\
Az$-$ & $1.45$ & $0.243$ & $0.0264$ & $3365$ & $2.57$ & $5808$ & $41.8$\\
\\ \hline
\end{tabular}
\end{table}

\begin{figure*}
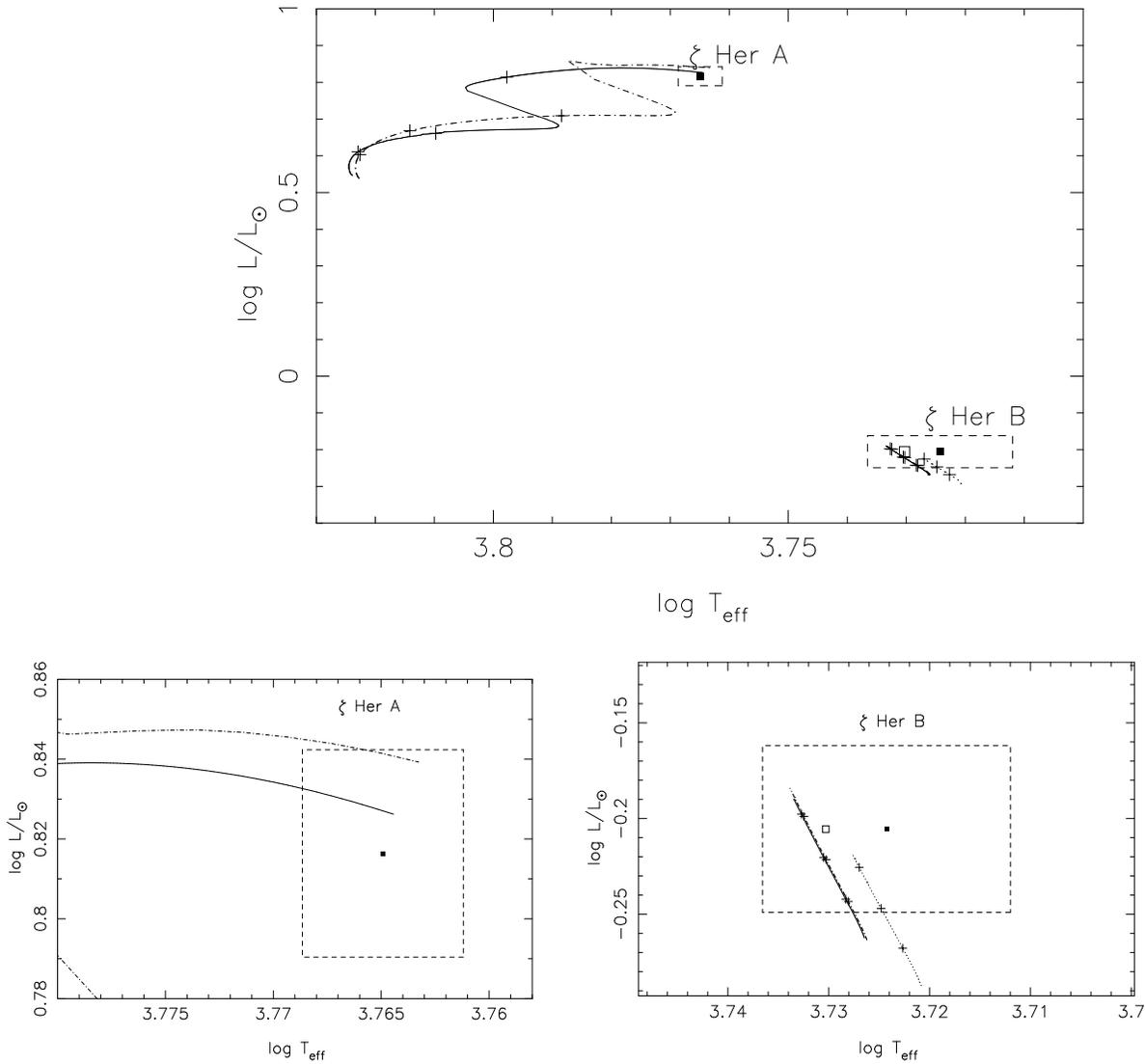

\vbox{
\centerline{
\psfig{figure=1660f1.eps,height=8.5cm,angle=270}
}
\vspace{0.5truecm}
\hbox{\psfig{figure=1660f2.eps,height=5.5cm,angle=270}
\hspace{0.5truecm}
\psfig{figure=1660f3.eps,height=5.5cm,angle=270}}
}
\caption{
Evolutionary tracks in the H-R diagram for models of $\zeta$\,Her\,A, \& B
without overshooting (full), with overshooting of $0.2\,H_{\rm p}$ for the 
convective core of $\zeta$\,Her\,A (dot-dash-dot)
and with a larger metallicity of 0.05 dex for $\zeta$\,Her\,B (dotted).
The open squares correspond to the locus of
$\zeta$\,Her\,B assuming this larger metallicity.
Dashed rectangles delimit the uncertainty domains.
Top panel: full tracks from {\sc zams}.
The ``+'' signs  denote 1\,Gyr time intervals along the evolutionary tracks.
Bottom left and right panels: enlargements around the observed
$\zeta$\,Her\,A \& B loci.
}\label{fig:AB}
\end{figure*}

\subsection{Seismological analysis of $\zeta$\,{\rm Her\,A}}\label{sec:seis}
The sub-giant star  $\zeta$\,Her\,A has a convective envelope
which may stochastically excite oscillations as is the case in the Sun.
Some seismic  observations of $\zeta$\,Her\,A have been done by
Marti\'c et al.~(\cite{mlsab01})
and seem to indicate a narrow excess of power around a maximum 
at $675\,\mu$Hz in the power spectrum.
It is thus interesting to consider the seismic properties of the model
selected by the calibration.

For our $\zeta$\,Her\,A model without overshooting, we have computed a 
set of adiabatic frequencies of the normal modes 
for degrees $\ell=0,1,2,3$ in the frequency range $300$ to $1\,000\,\mu$Hz.
The model corresponds to an evolved star which has burnt all the 
hydrogen core and presents a radiative core and a convective
envelope starting at $r_{\rm c} \sim 0.752\,R_\star$,
($M_{\rm c} \sim 0.9929\,M_\star$). At age $t_{\zeta\rm\,Her}=3\,387$\,Myr
the convective core which is present during the main sequence evolution
 has disappeared about $440$\,Myr ago. A zone of varying chemical 
gradient was formed between the outer edge of the initial convective 
core and the center.  This $\mu$-gradient gives a rapid 
variation of the sound speed and a large value of the maximum of
Brunt-V\"ais\"al\"a 
frequency $N$ (e.g. Unno et al.~\cite{uoass89}):
\[
N^2\equiv \frac gr\left(\frac1{\Gamma_1}\frac{\partial\ln P}{\partial\ln r}
-\frac{\partial\ln\rho}{\partial\ln r}\right),
\] 
of the order of $3\,500\,\mu$Hz --
as usual $P$ is the pressure, $\rho$ the density, $\Gamma_1$
the first adiabatic
exponent and $g$ the local gravity.
It acts like a potential well that may trap gravity
 waves. Figure~\ref{fig:prop} shows the profiles
 of the Brunt-V\"ais\"al\"a frequency $N$ and of Lamb frequencies $L_\ell$
($\ell=1,\,2,\,3$)
with respect to the normalized radius $r/R_\star$, viz.
the corresponding propagation diagram. Recall that the Lamb frequency writes:
\[
 L_\ell\equiv\frac{v_{\rm s}}r\sqrt{\ell(\ell+1)},
\]
eg. Unno et al.~(\cite{uoass89}), $v_{\rm s}$ is the sound velocity. 

Looking at the set of computed frequencies, we see that for each degree, 
the oscillation spectrum is no longer composed of two separated
sets of modes with acoustic ($p$-modes) and gravity ($g$-modes) behavior 
as in solar-like stars but it is a complicated superposition of these 
two sets. Some of these waves have a dual character as they  behave like
 pressure waves in the envelope of the star and gravity waves in the core. 
This greatly complicates the classification of the non-radial modes.

The distinction between the $g$- and $p$-modes
can be made by considering their normalized integrated kinetic energy
(or inertia) $E_{\rm n, \ell}$ (e.g. Christensen-Dalsgaard
\& Berthomieu~\cite{cb91}).
 The energy of the $p$-modes does not depend on the degree.
Figure~\ref {fig:ener} plots
this quantity as a function of the frequency for $\ell=0,1,2,3$.
The $g$-modes have
a much larger energy than the $p$-modes.
It appears  that for $\ell=1$ the modes all have both 
energy in the core and in the envelop, so they are mixed modes.
Figure~\ref{fig:fonc}, upper panel, shows the distribution of
kinetic energy density of two of these modes along the radius of the star.
 For $\ell=2,3$ we 
have a succession of modes with  energy either in the core or in the 
envelope with a few mixed modes as seen in Fig.~\ref{fig:fonc}, lower panel.

\begin{figure}
 \centerline{
 \psfig{figure=1660f4.eps,height=6.cm,angle=0}}
 \caption[]{Propagation diagram: profiles as functions of the nondimensionned
stellar radius, of Brunt-V\"ais\"al\"a
frequency $N$ and Lamb frequencies $L_{\ell}$ for degrees $\ell=1$ (dotted),
$\ell=2$ (dot-dash) and $\ell=3$ (dot-dot-dot-dash).
The horizontal lines correspond to the 
two modes $\ell=2,\ \nu=563.5\, \mu$Hz and $\nu=593.8\,\mu$Hz, the first one
with gravity behavior and the second one with acoustic behavior.
Figure~\ref{fig:fonc} shows their eigenfunctions. The full line indicates
the region of
the star where the modes are propagative.
}\label{fig:prop}
\end{figure}
\begin{figure}
 \centerline{
 \psfig{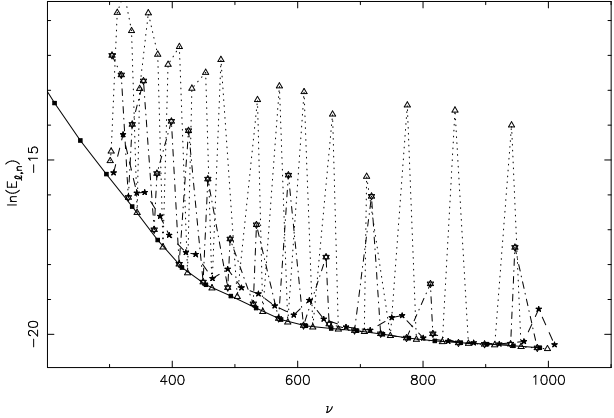}}
 \caption[]{ Logarithm of the normalized integrated kinetic energy
 $E_{\rm n,\ell}$ of the modes as a
 function of frequency (in $\mu$Hz) for modes of degree  $\ell=0$
 (full), $\ell=1$ (dashed), $\ell=2$ (dot-dash), $\ell=3$ (dotted). Note that
 all the points corresponding to $p$-modes are on the full line.
 }\label{fig:ener}
\end{figure}
\begin{figure*}
 \centerline{
 \psfig{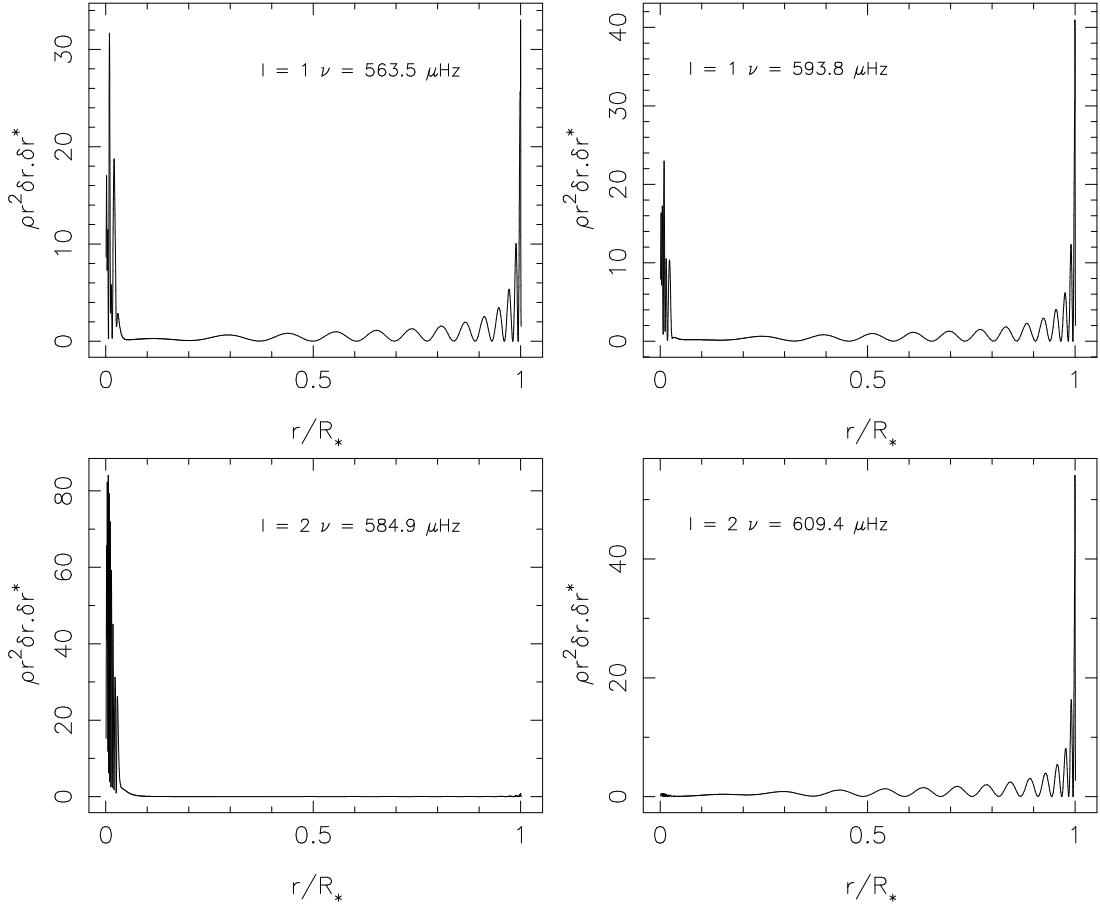}}
 \caption[]{
 Density of kinetic energy $\rho r^2\delta r.\delta r^\ast$ of two
 modes with consecutive radial orders for $\ell=1$ (upper panels)
 and $\ell=2$ (lower panels) as a function of the normalized radius
 ($\rho$ is the density and $\delta r$ the displacement).
 Note that the  modes $\ell=1$ are mixed modes while the modes $\ell=2$
 are alternatively $g$- and $p$-modes.
}\label{fig:fonc}
\end{figure*}
The set of frequencies with acoustic behavior can be analyzed
by classical asymptotic tools. The acoustic characteristic mean large
frequency spacing, also called ``large difference'',
corresponds to a mean of the quasi uniform spacing between
mode frequencies with the same 
degree and consecutive orders, is given by:
\begin{equation}\label{eq:dnu}
\overline{\Delta\nu_0}=\left(2\int_0^{R_{\rm s}}
\frac{{\rm d}r}{v_{\rm s}(r)}\right)^{-1}\sim 42\, \mu{\rm Hz};
\end{equation}
where $R_{\rm s}$ corresponds to the radius of the outermost shell of the
model.
The scaling of Kjeldsen \& Bedding~(\cite{kb95}) gives a close value:
\[
\overline{\Delta\nu_0}_{\rm sc}=\overline{\Delta\nu_0}_\odot\sqrt{\frac m{r^3}}
=39.7\,\mu{\rm Hz},
\]
where $m$ and $r$ are respectively the mass and the radius in solar units and
$\overline{\Delta\nu_0}_\odot=134.9\,\mu$Hz.
The computed
frequencies are fitted to the following asymptotic relation (Berthomieu et
al.~\cite{bpml93}):
\begin{equation}\label{eq:asy}
\nu_{n,\ell} = \nu_0 +\Delta\nu_\ell (n-n_0) +a_\ell(n-n_0)^2.
\end{equation}
With radial order $n$ between 10 and 20 and $n_0=15$
we obtain values for the
mean large difference at a given degree listed Table~\ref{tab:sep}.
They are close to the large difference $\overline{\Delta\nu_0}$, cf.
Eq.~(\ref{eq:dnu}), though smaller.
\begin{table}
\caption[]{
Theoretical global asymptotic quantities (in $\mu$Hz) describing
the $p$-mode oscillations according to Eq.~(\ref{eq:asy}) for degrees
$\ell=0,1,2,3$.
}\label{tab:sep}
\begin{tabular}{lllllllll} \\  \hline \\
  $\ell$        &$\nu_0$ &$\Delta\nu_\ell$ &$ a_\ell$ \\ 
\\ \hline \\
   0  &   654.0  &   40.7   &   0.130 \\
   1  &   659.5  &   38.8  &   0.336 \\
   2  &   650.4  &   40.8  &   0.094 \\
   3  &   645.0  &   40.7  &   0.111 \\
\\ \hline
\end{tabular}
\end{table}
These values are to be compared to the observations of
Marti\'c et al.~(\cite{mlsab01}) who derive a value around $42\,\mu$Hz
from the construction of an echelle diagram.
Figure~\ref{fig:ldif} plots the large differences
$\nu_{n,\ell}-\nu_{n-1,\ell}$ with respect to frequency.
Except for modes $\ell=1$, the points corresponding to high frequency modes
are close to the same flat curve around  $41\,\mu$Hz, 
with oscillations due to the rapid variation of the adiabatic exponent 
$\Gamma_1$ in the
helium ionization zone. As in the Sun, the $g$-modes have small amplitudes at
the surface and thus they will be hardly observable.
Therefore it is probable that the echelle diagram
used by the observers to extract the frequencies, taking 
account of their asymptotic distribution, will work for $\ell=0,\,2,\,3$ but
not for the mixed modes $\ell=1$.
The fit of the small differences:
\begin{eqnarray*} 
d_{02}(n)&=&\nu_{n,\ell=0}-\nu_{n-1,\ell=2}\sim\delta\nu_{02}+a_0(n-n_0),\\
d_{13}(n)&=&\nu_{n,\ell=1}-\nu_{n-1,\ell=3}\sim\delta\nu_{13}+a_1(n-n_0),
\end{eqnarray*}
gives the following results ($\mu$Hz):
\begin{eqnarray*}
\delta\nu_{02}&=&\ 3.963,\ a_0=-0.117,\\
\delta\nu_{13}&=&15.820,\ a_1=-1.774.
\end{eqnarray*}
Figure~\ref{fig:sdiff} shows the asymptotic behavior of
$\delta\nu_{02}$ with aligned points and
the large dispersion of points
for $\delta\nu_{13}$ due to the mixed character of the modes $\ell=1$.
\begin{figure}
 \centerline{
 \psfig{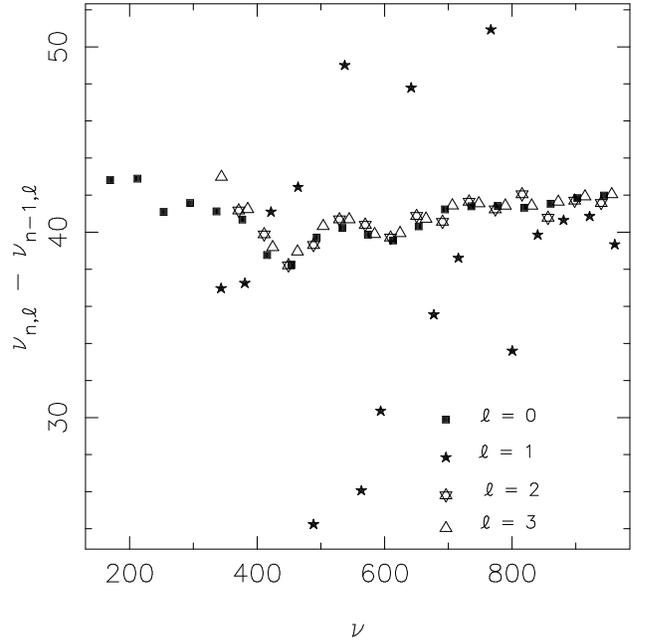}}
 \caption[]{Frequency differences between acoustic modes of
 same degree and consecutive radial order as function of the frequency
 (in $\mu$Hz).
 }\label{fig:ldif}
\end{figure}
\begin{figure}
 \centerline{
 \psfig{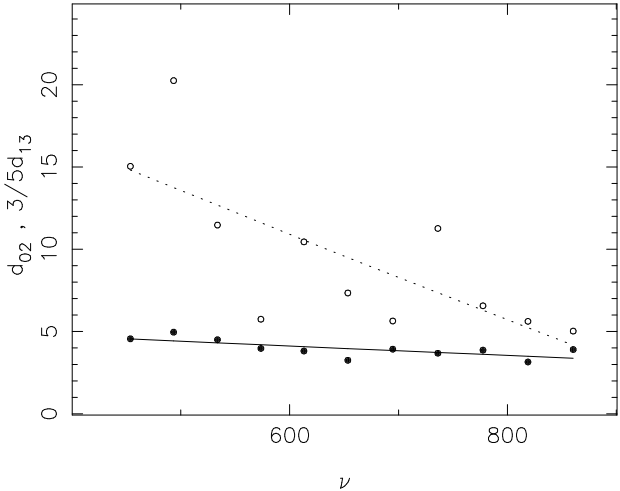}}
 \caption[]{Variation of the small differences $d_{02}$ (full dots)
and  $\frac35d_{13}$ (open dots),
 as a function of the frequency (in $\mu$Hz). The full and dotted lines
 represent their linear fit.}\label{fig:sdiff}
\end{figure}

\medskip
The calibrated model of $\zeta$\,Her\,A calculated with overshoot
of its convective core,
has a larger radius inducing a smaller value of $\overline{\Delta \nu_0}$
as seen in Table~\ref{tab:glob}.

\section{Discussion}\label{sec:dis}
The Achilles' heel of our calibration is the derivation of effective
temperature and luminosity of $\zeta$\,Her\,B. It is based on the only
reliable measurements of B and V magnitudes {\em of each component}
by {\sc tycho}
and {\sc hipparcos}.
The Bessell~(\cite{b00}) photometric calibration allows us to derive
the standard Johnson B and V magnitudes
either from $(\rm B_T \& V_T)$ {\sc tycho} magnitudes, or from
$(\rm B_T \& V_T)$ and $\rm H_P$ {\sc hipparcos} magnitudes.
In the case of $\zeta$\,Her\,B, B and V magnitude
values obtained in each way differ from each other by more than is
expected from the accuracy of measurements. Moreover, fixing the color index
(B-V), the effective temperatures calculated with different
photometric calibrations (Magain~\cite{m87},
Alonso et al.~\cite{aam96}, Flower~\cite{f96}) differ by more than 150\,K.
Fortunately, the scattering is not so large
with the bolometric correction $\rm BC_V$. As a result, the
effective temperature and the
luminosity of $\zeta$\,Her\,B are hardly known with a relative accuracy
better than 3\% and 10\% respectively. As a matter of comparison we have
derived the effective temperature and the
luminosity of $\zeta$\,Her\,A with better accuracy,
respectively 0.8\% and 6\%.

We are aware that our models suffer from the absence of 
diffusion of chemicals species. We do not introduce it as we do not have
a satisfactory description of the physical process which acts against 
 the too large efficiency of the gravitational settling in the envelope
of a main sequence stellar model with mass larger than $M\goa1.3M_\odot$
 i.e. without a significant outer convection zone (e.g. Schatzman~\cite{s69},
 Turcotte et al.~\cite{trmir98}).
The derived metallicity of $\zeta$\,Her\,A could not be
representative of
the initial mixture that formed $\zeta$\,Her\,A \& B. Because of the large
difference in
mass between the components, the diffusion has been more
efficient in $\zeta$\,Her\,A than in $\zeta$\,Her\,B
and therefore the adopted metallicity for $\zeta$\,Her\,B needs to be increased.
As the mass of $\zeta$\,Her\,B is close to the solar one, we can safely assume
that the change in metallicity has occurred at the same rate in the Sun and in 
$\zeta$\,Her\,B. From {\sc zams} to the age
$t\approx3\,400$\,Myr, the metallicity of the Sun is increased by about $0.05$\,dex.
Using Magain's~(\cite{m87})
formula, an $\rm [Fe/H]_B$ increase corresponds to
an increase of $\sim+75$\,K in effective temperature.
 Figure~\ref{fig:AB} shows that the locus (open square)
 of $\zeta$\,Her\,B in the {\sc hr} diagram with $T_{\rm eff\,B}$ larger by $75$\,K
 is closer than before to the evolutionary tracks. To go further one needs
 a measurement of the metallicity of $\zeta$\,Her\,B.

With $m_{\rm A}=1.50\,M_\odot$ and $m_{\rm B}=1.00\,M_\odot$
it was not possible to obtain simultaneous satisfactory adjustments,
within the error boxes, for both components. 
Realistic solutions are found with $m_{\rm A}=1.45\,M_\odot$. That may
indicates that the suspected duplicity of $\zeta$\,Her\,A is perhaps real.
In such a case the hypothetical unseen
component $\zeta$\,Her\,a will be less massive than previously announced,
$m_{\zeta\rm\,Her\,a}\loa0.05\,M_\odot$ 
and will be a brown dwarf or a giant planet.
Though Table~\ref{tab:glob} exhibits insignificant small
differences between mixing length parameters of $\zeta$\,Her\,A \& B,
we obtained values of the order of unity,
as expected with the Canuto \& Mazitelli~(\cite{cm91}, \cite{cm92})
convection theory.
The differences between the C95
and our calibrations mainly result from the difference in distance.
In the {\sc hr} diagram the locus of our $\zeta$\,Her\,A model is found
in the Hertzsprung gap soon
after the main sequence, as expected for a sub-giant star.
In this part of the {\sc hr} diagram, the effective temperature of a star model
varies rapidly with respect to time.
So the age is very sensitive to small changes of physics or of
modeling parameter. The small uncertainty we give for the
age corresponds to the crossing time of the error box in effective temperature.
 It is valid only
with the physics we used and with the central values of other
modeling parameters.
 In their study C95 obtained a value for the age:
$t_{\zeta\,\rm Her}=4.0\pm0.4$\,Gyr close to ours. We emphasized the fact that
the crossing time of the C95 error box
in effective temperature ($\pm40$\,K) is about twenty times smaller
than their estimated accuracy in age.

Table~\ref{tab:acc} lists the ages, the radii, the effective temperatures
and the large differences of $\zeta$\,Her\,A models
computed with modeling parameters within their error bars, but for mixing-length
parameters. The age of the model Ag$-$ ($respt.$ Ag$+$) 
corresponds to the time elapsed from {\sc zams}
to the instant where the effective temperature crosses the left
($respt.$ right) limit of the error box in the {\sc hr} diagram, namely
$\log T_{\rm eff}=3.768$ ($respt.$ $\log T_{\rm eff}=3.761$).
The ages of other models listed in Table~\ref{tab:acc} corresponds
to the time elapsed from {\sc zams}
to the instant where the effective temperature crosses the central value
$\log T_{\rm eff}=3.765$. The variations of $\overline{\Delta\nu_0}$
reflect the differences in radius and mass of star models. 
Table~\ref{tab:acc} shows that one derives
unrealistic precision for the modeling parameters
by defining their accuracy in
such a way that any combination of them, within their error bars,
will provide models of $\zeta$\,Her\,A \& B within
the observational constraints.
Owing to the weakness of the observing material we do not attempt to derive the
accuracy of the modelling parameters using the method developed by Brown et
al.~(\cite{bcwg94}).
A more sensitive modeling parameter
is the age due to the fast post main-sequence evolution of $\zeta$\,Her\,A. 
The accuracies of modeling parameters listed in
Table~{\ref{tab:glob} may be optimistic. They mean that,
for a value of any modeling
parameter within its
interval of accuracy, {\em one can find} a set of other modeling
parameters, each of them within its own accuracy limit, in such a way that they
will provide models of $\zeta$\,Her\,A \& B within
the observational constraints.

\section{Conclusion}\label{sec:con}
Detailed evolutionary calculations of the visual binary system
$\zeta$\,Herculis have
been performed using {\sc opal} opacities and equation of state, {\sc nacre}
thermonuclear reaction rates, Canuto \& Mazitelli~(\cite{cm91},~\cite{cm92})
convection theory.
The effective temperature, surface gravity and 
metallicity of $\zeta$\,Her\,A have been reestimated
using published spectroscopic data 
(Chmielewski et al.~\cite{cccls95}, Th\'evenin~\cite{t90})
and new photometric data of the {\sc tycho}
catalogue (Fabricius \& Makarov~\cite{fm00}).
The luminosity of $\zeta$\,Her\,A and the effective temperature and luminosity
of $\zeta$\,Her\,B are calculated according to {\sc tycho}
and {\sc hipparcos} photometric data
and Bessell~(\cite{b00}) calibrations. The sum of masses have been estimated
using updated astrometric relative orbit and parallax determinations.
The mass fraction is derived from spectroscopic and astrometric
orbits. We have determined the most reliable solution 
within the confidence domains of the
observable constraints via a $\chi^2$ minimization.
Each solution is characterized by
$\wp=\{t_{\rm\zeta\,Her},m_{\rm A},m_{\rm B},Y_{\rm i},
{[\rm\frac{Fe}H]_i},\Lambda_{\rm A},\Lambda_{\rm B}\}$,
where $t_{\rm\zeta\,Her}$ is the
 age of the system, $m_{\rm A}$ and $m_{\rm B}$ the masses of components A \&
 B respectively, $Y_{\rm i}$ the initial helium content,
 $[{\rm\frac{Fe}H]_i}$ the initial metallicity and 
 $\Lambda_{\rm A}$ and $\Lambda_{\rm B}$ the convection parameter
 of each star model.
We obtained $\wp_{\zeta\,\rm Her}=
\{3\,387\,{\rm Myr},1.45\,M_\odot,0.98\,M_\odot, 0.243,0.0269,0.92,0.90\}$. As a
sub-giant, in the
{\sc hr} diagram $\zeta$\,\rm Her\,A is located beyond the main-sequence
in the ``Hertzsprung gap'', where the star loci move rapidly.
So, all other modeling parameters being fixed, the locus of $\zeta$\,\rm Her\,A
takes only a couple of ten Myr to cross over the error box in effective temperature. 
Therefore, {\em fixing the physics}, the modeling parameters
of the system appear artificially well defined.
Fixing the masses obtained previously, we have calibrated
the binary system assuming
an overshoot of $0.20\,H_{\rm p}$, for the convective core
of $\zeta$\,Her\,A. We did not obtain significantly different
modeling parameters but a larger value for the age, namely $\sim+250$Myr.

The adiabatic oscillation spectrum of $\zeta$\,Her\,A is found to be a
complicated superposition of
acoustic and gravity modes. Some of these waves  have a dual character.
This greatly complicates the classification of the non-radial modes.
For $\ell=1$ the modes are mixed modes with both 
energy in the core and in the envelope, they are mixed modes.
For $\ell=2,3$ 
there is a succession of modes with energy either in the core or in the 
envelope with a few mixed modes. This will have an implication for
the properties of the echelle diagram used by observers which will have a smooth
behavior only for the modes $\ell=0,\,2,\,3$. 
 The large difference
is found to be of the order of $\Delta\nu_0\approx 41\,\mu$Hz close to the
preliminary value derived from observations by Marti\'c et al.~(\cite{mlsab01}).

Six years ago, in the conclusion of their study,
 Chmielewski et al.~(\cite{cccls95}) requested
{\it ``[..] to go further, (i) a better photometry of component B and (ii)
the exact value of the parallax''}.
Our work shows that the calibration of $\zeta$\,Her remains a difficult task,
 even with the disposal of improved observational data.
Among the binaries to be calibrated with some confidence,
$\zeta$\,Herculis is one of the most interesting owing to the difference of
evolutionary state of components resulting from their mass difference.
Though difficult, $\zeta$\,Her is a reachable target for
modern spectroscopic and photometric apparatus.  $\zeta$\,Her
deserves interest to improve the accuracy of the modeling constraints hence,
 of stellar models and seismological analysis. It is all the more desirable because
 seismic observations of $\zeta$\,Her\,A will hopefully give new  constraints.

\begin{acknowledgements}
We thank M.Marti\'c for having directed our attention to the opportunity to
revisit the calibration of the $\zeta$\,Her binary system and to undertake the
seismological analysis of the brightest component.
We would like to express our thanks to the unknown referee, 
for helpful advises.  
This research has made use of the Simbad data base, operated at
CDS, Strasbourg, France and of the WDS data base operating at USNO,
Washington, DC USA.
This work has been performed using the computing facilities 
provided by the OCA program
``Simulations Interactives et Visualisation en Astronomie et M\'ecanique 
(SIVAM)''.
\end{acknowledgements}

\end{document}